\documentclass[a4paper]{jpconf}
\usepackage{graphicx}
\begin{document}
\title{The evolution of wealth transmission in human populations: a stochastic model}

\author{G. Augustins$^1$, L. Etienne$^2$,  J-B. Ferdy$^3$, R. Ferrer$^2$,
B. Godelle$^2$, E. Pitard$^4$, F. Rousset$^2$}

\address{$^1$ Laboratoire d'Ethnologie et de sociologie comparative, Universit\'e Paris X}
\address{$^2$ Institut des Sciences de l'Evolution de Montpellier, Universit\'e Montpellier 2}
\address{$^3$ Laboratoire Evolution et Diversit\'e Biologique, Universit\'e Toulouse 3}
\address{$^4$ Laboratoire de Physique Charles Coulomb, Universit\'e Montpellier 2-CNRS}



\ead{estelle.pitard@univ-montp2.fr}

\begin{abstract}
Reproductive success and survival are influenced by wealth in human populations.
 Wealth is transmitted to offsprings and  strategies of transmission vary over time  and among populations, the main variation
 being how equally wealth is transmitted to  children.
Here we propose a model where we simulate both the dynamics of wealth in a population and the evolution of a trait that determines how wealth is transmitted from parents to offspring, in a darwinian context.
\end{abstract}

\section{Introduction}

 

 
 



It has been shown that in Eurasian societies, there exists a link between family types and ideological values, such as individualism, liberalism or equality \cite{Todd}.
In this paper we are particularly interested in the way wealth is transmitted through generations in human populations. One is interested for example to the case of primogeniture (or unigeniture), i.e when the wealth is given only to one preferred child, and this choice would be preferred to a case of more equal treatment i.e when the wealth is distributed evenly between children \cite{Borgerhoff}. These conditions are numerous, and can be for example cultural, historical or economical factors.

Here we are interested in the economical factors that influence this choice. Indeed, it is known that wealth has an influence on demography, and more precisely on fecundity. It also has an influence on the social status, which comprises the probability to marry, and hence, to have  descendants. It is also known that wealth of the parents influence the financial status of their children: statistically, the richer the parents, the richer the children \cite{Voland, Hopcroft}.
These facts are documented by several case studies, which can be in some cases very complex to interpret.

In the case of rural, traditional populations, the above facts seem to be less complex, and tend to show that the probability to reproduce increases as wealth increases. In these populations, wealth is measured in units of land, as money was not necessarily an indicator for richness, nor was even needed.

Concerning the wealth transmission rule in human populations, historical studies show that they are very variable, and depend on the population considered \cite{Hrdy}.
For example, in rural regions in France before industrialization, the rules are diverse:
egalitarian in Limousion but primogeniture in the Basque region.
 They are also variable in time. Take for example the population of settlers in New England in the 17th century: their rule was first egalitarian, having a large amount of land available. But after several generations, land becoming scarce, they changed the rule and chose primogeniture, as 
Alexis de Toqueville reports in his book "Democracy in America" (1835): "In Massachusetts, estates are very rarely divided; the eldest son generally takes the land, and the others go to seek their fortune in the wilderness. The law has abolished the right of primogeniture, but circumstances have concurred to re-establish it".


Our aim is to derive a simple model that allows to draw a link between the wealth distribution inside a population and how wealth transmission rule evolves.
We will start by constructing an economical model based on specific rules in a traditional, agricultural, closed population, that allows to quantify economical inequalities.
Then we choose a reproduction model, where reproduction is conditioned on wealth.
The transmission (or devolution) model consists in introducing a parameter $\rho$ which is the part given to the first, preferred, offspring.
The last ingredient is the mutation rate, which allows $\rho$ to vary randomly over the generations. This parameter is also subject to selection, through the value of the parent's wealth. 
The evolution of 
$\rho$ is considered and results in two distinct regimes.
In the case of strong economical inequalities, equal transmission of wealth is preferred, 
whereas in the case of less inequalities, primogeniture is preferred.

\section{An economical model of wealth based on land}

Here we focus on wealth distribution of closed, rural societies.
 In these agricultural societies, wealth $w$ is the amount of land owned, on which crops can be grown.
  The model is described as follows.

At time $t$, each individual $i$ owns some land ($=$ wealth) $w_i(t)$.
The total land of the community is supposed to be conserved over time: if there are $N$ individuals, then $w_{tot}=\sum_{i=1}^N w_i(t)$ is a constant.

 The production of crops -typically every year- results in a quantitative harvest $x_i(t)$, which is subject to some environmental noise of variance per unit of land $\sigma^2$, depending on climate variations, quality of soil and other potential factors.   It is also assumed that the harvest can be measured in units of land (actually it may be thought that both land and harvest can be measured in the same unit type, such as the number of seeds needed, then harvested after the plant has grown). For simplicity, we also assume that the productivity is equal to $1$, so that the harvest is finally equal on average -over the noise- to $w_i(t)$.
This can be written as: 
  $$x_i(t)=w_i(t) +\sqrt{\sigma^2 w_i(t)}  \eta_i(t).$$

 \noindent Here  $\eta_i(t)$ is a gaussian noise of zero mean and variance $1$, uncorrelated in time.

It has been observed that such communities choose a tax system, in a way to  decrease
variations of wealth between individuals. In such communities with little interaction with some outside world, money was not the correct tool for transactions. A certain amount of wealth -for example, seeds- is taken to each individual proportionally to his harvest, with rate   
  $\alpha$, and redistributed equally among all members of the community.
The amount of  harvest remaining after tax is:
  $x'_i(t)=(1-\alpha) x_i(t) + \frac{\alpha}{N} \sum_{i=1}^N x_i(t)$

 For each land owner, there is also  a maintenance cost for the land, corresponding for example to the
cost of buying the correct amount of seeds needed for the next crop. We will again estimate this quantity in the same unit as the wealth and the harvest, and assume that it is simply equal to
$w_i(t)$. Finally, what remains of the harvest after tax and maintenance cost is:
  $h_i(t)=x'_i(t) -w_i(t)$.

The quantity $h_i(t)$ can actually be positive or negative: 
 by chance, some individuals produce more, while others are not able to maintain their land. One can check that the average over the noise of the quantity $\sum_{i=1}^N h_i(t)$ is equal to zero for all times.
A way to erase debts and assure sustainability of the whole community is a redistribution of land among individuals. An individual with positive $h_i(t)$  can  convert it into buying land, while an individual with negative $h_i(t)$  can sell some land to compensate for bad harvest. In all cases, one can write the wealth at time $t+1$ as:
$w_i(t+1)=w_i(t)+ h_i(t).$

 
 
After such exchanges,  the land is redistributed, but with time, some stationary probability distribution  $P(w)$ is reached. Typically such an economical cycle (with a crop each year) lasts  a generation, namely approximately 25 years.



 
  
   
  
  
   

One can use the preceding equations to write a continuous time evolution of wealth $w(t)$:

  $$\frac{\partial w}{\partial t}=\alpha (w_0 - w(t)) +(1-\alpha) \sigma \sqrt{w(t)} \eta(t)$$

 \noindent with
  $w_0=w_{tot}/N$.
  
 From this Langevin-type equation of evolution, one can derive, in the Stratanovich sense \cite{vanKampen}, as we have here an extrinsic noise- the corresponding   Fokker-Planck equation for $P(w,t)$:

  $$\frac{\partial P(w,t)}{\partial t}= 
   - \frac{\partial}{\partial w}[(\alpha(w_0-w) + \frac{(1-\alpha)^2 \sigma^2}{4}) P(w,t)] 
   +\frac{1}{2} \frac{\partial^2}{\partial w^2}[(1-\alpha)^2 \sigma^2 w P(w,t)]$$
  
   The stationary solution comes directly as:
  $$P_{st}(w) \propto \frac{1}{w^{1-\mu}} \exp[-(\mu -\frac{1}{2}) \frac{w}{w_0}]$$

\noindent   with
  $\mu -\frac{1}{2}=\frac{2\alpha w_0}{(1-\alpha)^2 \sigma^2}$.

The resulting distribution functions have an exponential tail, except for a situation with no tax and no redistribution ($\alpha=0$) where the distribution of wealth is a power law. At fixed tax rate $\alpha$ (Figures 1 and 2), the distribution gets broader as the environmental variance increases ($\sigma^2 \rightarrow \infty$ also leads to a power law). If the environmental variance $\sigma^2$ is fixed (results not shown), reducing the tax rate $\alpha$ makes the distribution broader as well, inequalities between individuals increase.


Note that other economical models are studied to the literature, and mainly lead to power-law type of wealth distribution, such as in 
\cite{BouchaudMezard}. In this last study, the economical context is different, individuals exchange wealth through trading, and the stochastic noise describes the growth or decrease of wealth due to investment in stock markets. This noise term is then of the form $w_i(t)\eta_i(t)$ -instead of $\sqrt{w_i(t)}\eta_i(t)$ in our case-, to insure invariance of the evolution equation under a change of monetary units.







\begin{figure}[h]
\begin{minipage}{18pc}
\includegraphics[width=18pc]{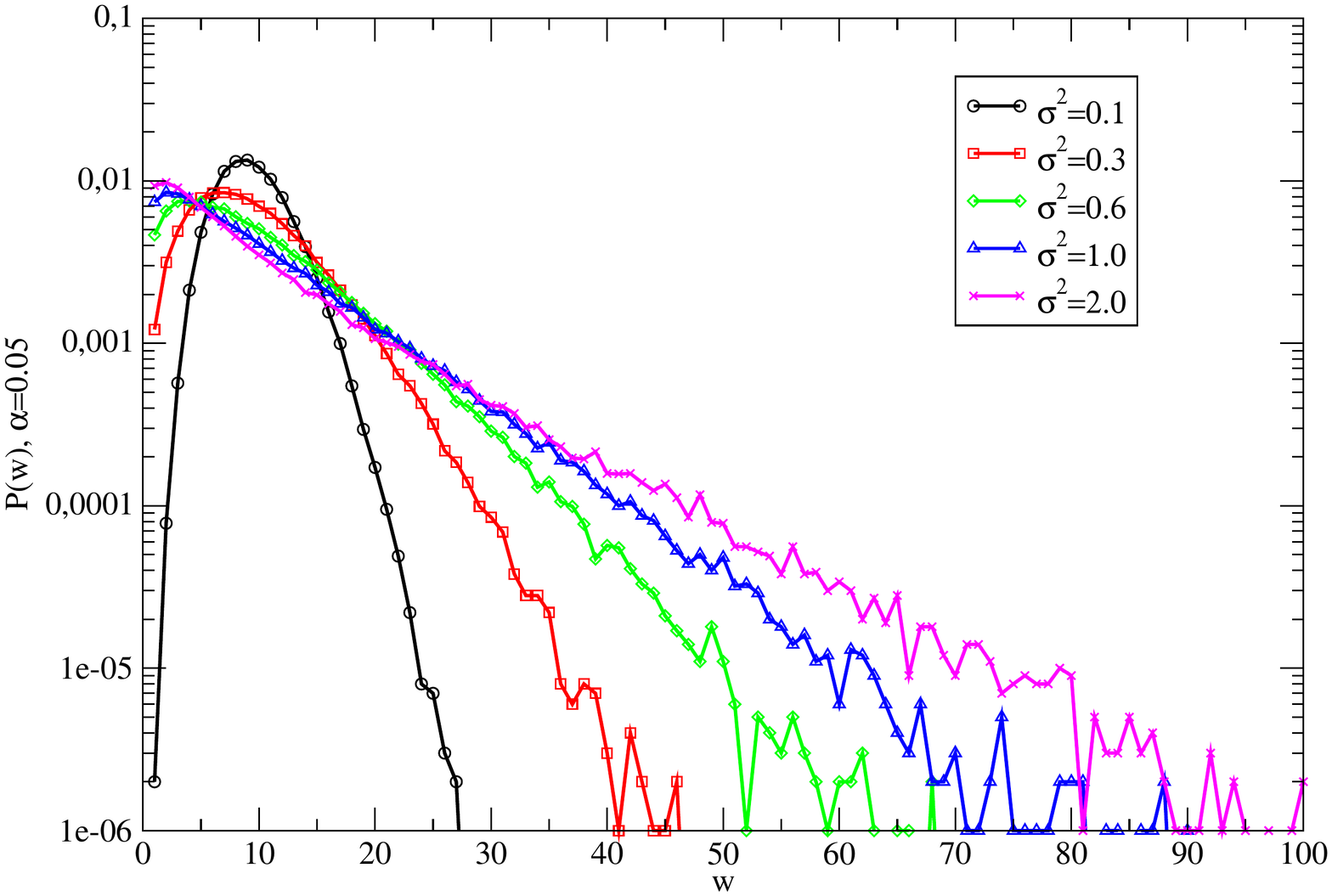}
\caption{\label{label}Simulation results for $P(w)$ with fixed tax rate, $\alpha=0.05$.
Inequalities increase when
 environmental variance $\sigma^2$ increases (from left to right).}
\end{minipage}\hspace{2pc}%
\begin{minipage}{18pc}
\includegraphics[width=18pc]{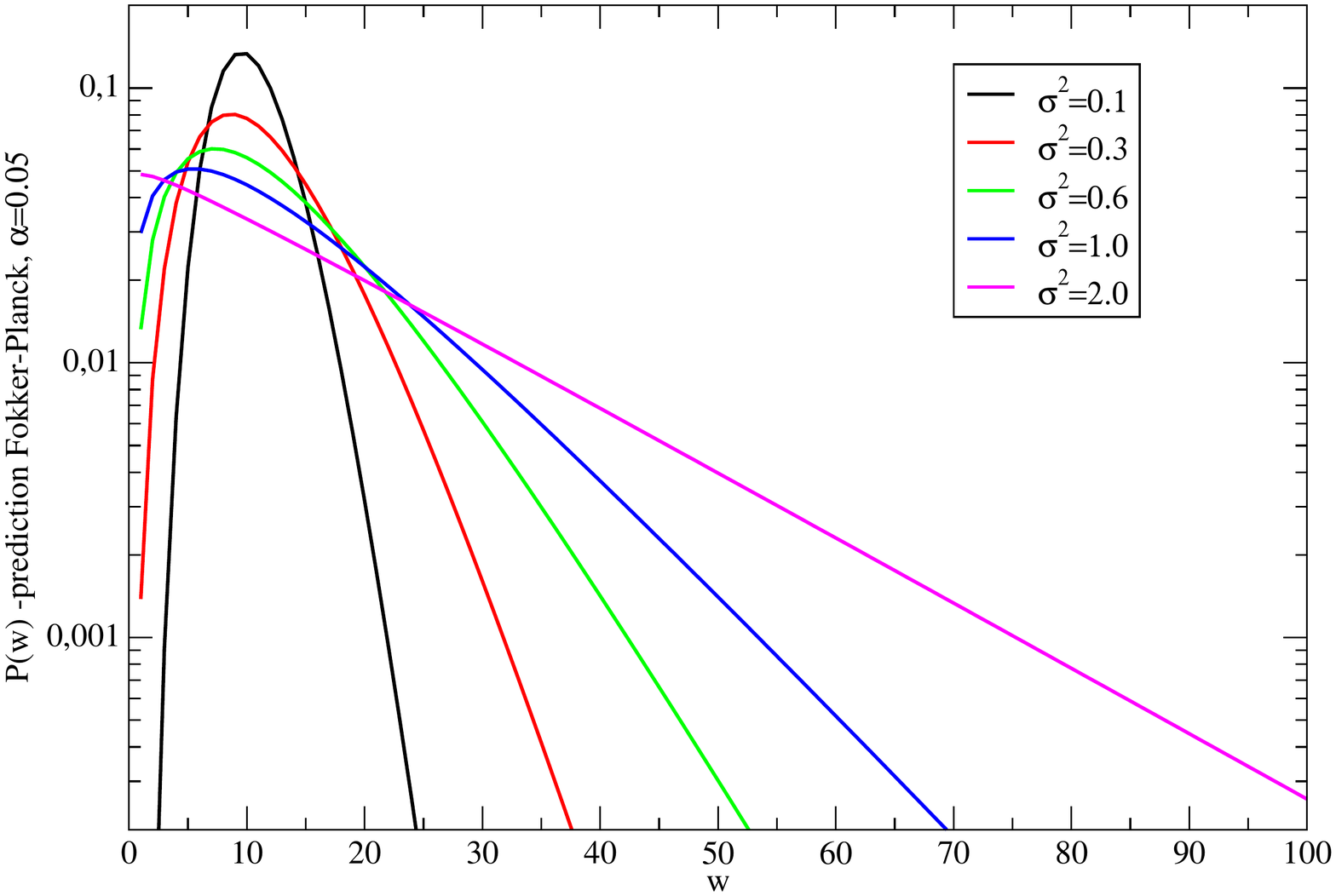}
\caption{\label{label}Theoretical results for $P(w)$ with the same parameters as in Figure 1.}
\end{minipage} 
\end{figure}




\section{Evolution model for the transmission of wealth}

Given the economical model described above, we are now interested in the way individuals transmit wealth to their offsprings.
We used a numerical simulation scheme, that follows over time the reproduction of individuals and the transmission of wealth to the children, for many generations. The scheme goes as follows:

{\it (i)} the economical cycle (described above) first runs over 25 years, namely a generation.

{\it (ii)} then reproduction occurs: an individual has a probability $p(w)$ to have children. $p(w)$ is an increasing function of $w$, typically a sigmoid going from 0 to 1 as $w$ varies from 0 to $\infty$. At this stage, wealth acts as a selection pressure for reproduction.

{\it (iii)} if there is reproduction, then the number of chidren $n$ is drawn from a Poisson distribution with mean $f$, which is the fecundity parameter.

{\it (iv)} during the devolution step, the father dies  (non-overlapping generations) and transmits all his wealth to his children. We choose the following rule of transmission: the fraction of wealth given to child $j$ is:

$$h_j=\rho w \frac{(1-\rho)^{j-1}}{1-(1-\rho)^n}, j=1,..n.$$

   \noindent Here the parameter $\rho$ is the fraction of father's land ($w$) given to the firstborn (preferred) child.
If {$\rho=1$} the firstborn child gets all the father's land (primogeniture).
If {$\rho=0$} all children receive the same fraction of the father's land: $h_j=w/n$, the transmission is egalitarian.
In the case where there is no offspring, wealth is redistributed among the population so as to preserve wealth ratios between individuals.
Then the economical cycle starts again for 25 years, and the whole process continues.
      
It has been argued \cite{Hrdy} that the type of  wealth transmission (modeled  here by $\rho$) is a trait subject to evolution, in the darwinian sense. In the model, it is a heritable trait. However, because this trait is expected to adjust to the wealth distribution, we let it evolve during the simulation:

      $$\rho_{child}=\rho_{father}(1+s)$$

The parameter $s$ is a mutation rate, drawn from a gaussian distribution of zero mean and small variance (typically $10^{-3}$): on average children transmit their wealth just as their father did, but 
       mutation creates random changes between father's and children's $\rho$.
 As the simulation goes on, starting from an initial value equal to $\rho=1/2$ for all individuals, we measured the final, stationary value of $\rho$ averaged over all individuals.



\begin{figure}
\begin{center}
\includegraphics[width=18pc]{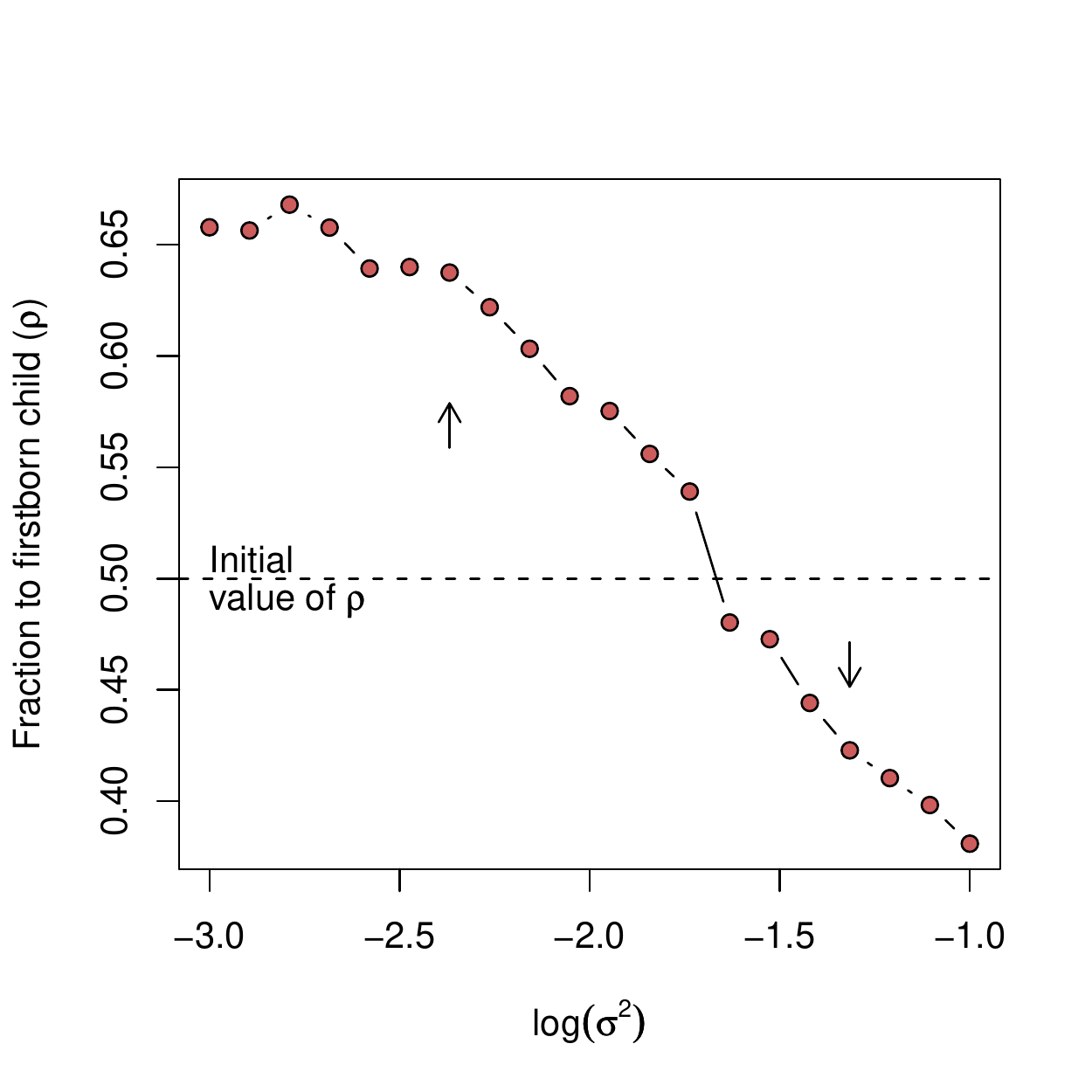}
\caption{\label{label}Evolution of the fraction given to first child $\rho$, after many generations, for different values of the environmental noise $\sigma^2$. Here, 50 independent simulations were run 
for 50000 generations and $\alpha=0.04, f=1.75, N=10^3$.}
\end{center}
\end{figure}

The results (Figure 3) reveal the existence of two regimes of selection of $\rho$.
When the environmental variance is weak,
 the fraction $\rho$ given to the firstborn child increases, which means that primogeniture emerges as the rule of transmission.
When environmental variance is high,
  the fraction $\rho$ given to the firstborn child decreases and one tends to the regime of egalitarian transmission.
 These features can actually be well understood in the framework of our model, and rely much on the basic assumption that wealth is a condition for successful reproduction.



 In the case of   narrow $P(w)$ ($\sigma^2$ small or $\alpha$ large),
many individuals are moderately rich or poor.
 If $\rho$ is too small, many offsprings will not reproduce 
because they are too poor. Because
  $\rho$ is determined by the survivors, it evolves to larger values and leads to favor primogeniture. 
 In the case of   broad $P(w)$ ($\sigma^2$ large or $\alpha$ small),
 the population is composed of many poor people but  there is also a significant fraction of very  rich people.
For the rich, if $\rho$ is  too large, less offsprings will reproduce than if $\rho$ is small because even if this case, they are sufficiently rich.
  $\rho$ is better small for the rich, so  on average it evolves to smaller values and  more egalitarian transmission.

\section{Conclusion}

Our results show that economical systems that decrease inequalities promote primogeniture. On the contrary, economical systems that increase inequalities promote more egalitarian transmission rule, which is  a priori counter-intuitive.
This last result is new and we do not expect it to depend on the type of trading (it is a priori sufficient that the wealth distribution is broad and close to a power-law). However, if the value of $\rho$ is more egalitarian on average, it can hide strong disparities between the poor and the rich. This could lead to further work, in particular in order to understand how wealth is redistributed -or not- among families.
Moreover, comparison with demographic data will be needed \cite{Todd}.
Finally, one can imagine a more general model, where the fecundity could also be an evolving parameter: this could help to understand  the 
  conditions for a demographic transition in time.




\smallskip


\section*{References}
\medskip
\begin{thebibliography}{9}


\bibitem{Todd} Todd E 2013 {\it L'origine des syst\`emes familiaux}, Gallimard. Todd E 1999 {\it La diversit\'e du monde: famille et modernit\'e}, Seuil.

\bibitem{Borgerhoff} Borgerhoff Mulder M and al. 2009 {\it Science} {\bf 326} 682.

\bibitem{Voland}Voland E 1990 {\it Behav Ecol Sociobiol} {\bf 26} 65.

\bibitem{Hopcroft} Hopcroft R 2006 {\it Evolution and Human Behavior} {\bf 27} 104.


\bibitem{Hrdy} Hrdy S and Judge D 1993 {\it Human Nature}  {\bf 4} 1.

\bibitem{vanKampen} van Kampen NG 1981 {\it J  Stat  Phys} {\bf 24} 175.

\bibitem{BouchaudMezard} Bouchaud J-P and M\'ezard M  2000 {\it Physica} A {\bf 282}  536.


\end{thebibliography}
\end{document}